\documentclass[twocolumn,showpacs,preprintnumbers,amsmath,amssymb]{revtex4}

 \usepackage{graphicx}
 \usepackage{amsmath}
 \usepackage{amsfonts}
 \usepackage{amssymb}
 \usepackage{pstricks}
 \usepackage{psfrag}
 \usepackage{epsfig}
 \usepackage[ansinew]{inputenc}

 \newcommand{\vecd}[1]{\mathbf{#1}}
 
 \newcommand{\tens}[1]{\sf {#1}}
 \newcommand{\mat}[1]{\tens{#1}}

\begin{document}

\title{Direct measurement of shear-induced cross-correlations of Brownian motion}

\author{A.~Ziehl$^{1}$, J.~Bammert$^{2}$, L.~Holzer$^{2}$, C.~Wagner$^{1}$, W.~Zimmermann$^{2}$}

\affiliation{$^1$ Technische Physik, Universit\"at des Saarlandes, 66041 Saarbr\"ucken, Germany\\ 
$^2$ Theoretische Physik I, Universit\"at Bayreuth, 95440 Bayreuth, Germany }

\date{September 2, 2009}

\begin{abstract}
Shear-induced cross-correlations  of particle fluctuations perpendicular and along stream-lines are investigated experimentally and theoretically.
Direct measurements of the Brownian motion of micron-sized beads, held by optical tweezers in a shear-flow cell, show a strong time-asymmetry in the cross-correlation, which is caused by the  non-normal amplification of fluctuations.
Complementary measurements on the single particle probability distribution substantiate this behavior and both results are consistent with a Langevin model. 
In addition, a shear-induced anti-correlation between orthogonal random-displacements  of two trapped and hydrodynamically interacting particles is detected, having one or two extrema in time, depending on the  positions of the particles. 
\end{abstract}


\maketitle

The Brownian motion of particles in fluids and their hydrodynamic interactions
are of central importance 
in chemical and biological physics as well as in material science and engineering
\cite{Einstein:1905.1,Dhont:96,Stone:2001.1,Ottino:2004.1}. However
our understanding of the dynamics of particles in flows
is still far from complete.
Direct observations of particles 
at the mesoscale substantially 
contribute to our understanding
of their dynamics. At this scale optical tweezers are
a powerful experimental technique \cite{Tweezers} with a number
of innovative applications. They include the detection of
anti-correlations between hydrodynamically interacting Brownian particles \cite{Quake:99.1}, propagation of hydrodynamic interactions \cite{Bartlett:2002.1},
short-time inertial response of viscoelastic fluids \cite{SchmidtChr:2005.1},
two-point microrheology \cite{WeitzD:2000.1}, 
anomalous vibrational dispersion \cite{Quake:2006.1} and particle sorting
techniques \cite{Grier:2002.1}.

Neutral colloidal particles moving relatively to each other interact via
the fluid and these hydrodynamic
interactions decay with the particle distance \cite{Dhont:96}. 
In shear flow little is known 
about the dynamics of
Brownian particle motion and the hydrodynamic interaction effects
in spite of their fundamental relevance and importance in applications
in microfluidics,  
Taylor dispersion \cite{TaylorGI:1953.1} 
and in fluid mixing \cite{Ottino:2004.1,Steinberg:01.2}. 
In time dependent fields and in shear 
flow surprising deterministic  particle dynamics
may be induced by hydrodynamic interactions \cite{Holzer:2006.1}.
For polymers it is the interplay of shear flow and fluctuations
which leads already at low Reynolds numbers, to rich dynamics \cite{Chu:97.1}, 
the so-called molecular individualism \cite{deGennes:97.1}, causing
elastic turbulence even  in diluted polymer solutions 
\cite{Steinberg:00.1} and spectacular mixing behavior \cite{Steinberg:01.2}.

It is the contribution $({\bf u} \cdot \nabla){\bf u}$ 
to the Navier-Stokes  equation which
causes  interesting transient phenomena in shear flows
near the onset
of turbulence \cite{SGrossmann:2000.1}, as well as 
amplifications of fluctuations and their cross-correlations along
and perpendicular to straight streamlines
 \cite{Eckhardt:2003.1,Oberlack:2006.1}. 
A cross-correlation is also expected 
between orthogonal particle-fluctuations
in the shear plane, because random jumps of a particle
between neighboring  streamlines of different velocity
lead to a change of the particle's velocity 
and displacement  along the streamlines, similar as via fluctuations.
  In  some parameter ranges,  inertia effects may become
important \cite{Bedeaux:1995.1,MacKintosh:2005.1,Reeks:2005.1}. 
Cross-correlations between perpendicular fluid-velocity fluctuations
and perpendicular fluctuations of particles
are expected to be strongly asymmetric in time
  \cite{Eckhardt:2003.1,Reeks:2005.1,HolzBamm}.
In dynamic light-scattering experiments certain 
aspects of these shear-induced cross-correlations 
were observed
indirectly \cite{vandeWater:1998.1}, but
a direct measurement and characterization of related 
particle fluctuations is  missing.

Here we investigate in a linear shear flow
the fluctuations 
of a single particle 
in a potential minimum and 
of two hydrodynamically
interacting particles trapped by two neighboring potentials.
We use a special 
shear-flow cell, where  
one or two micron sized beads are held
at its center by  optical tweezers. The time-asymmetry of shear-induced 
cross-correlations were determined 
directly by measuring the particle's positional 
fluctuations.
 In addition
the probability distribution of a single 
particle in a trap was measured,
which can be also calculated in terms
of a Langevin model, similar to the correlations. Both the  
probability distribution and the correlation can
be fitted by using the same value of the
shear rate, which altogether gives 
a consistent picture of not yet directly observed
shear-induced cross-correlations of particle fluctuations.

\begin{figure}[htb]
\vspace{-2mm}
\epsfig{file=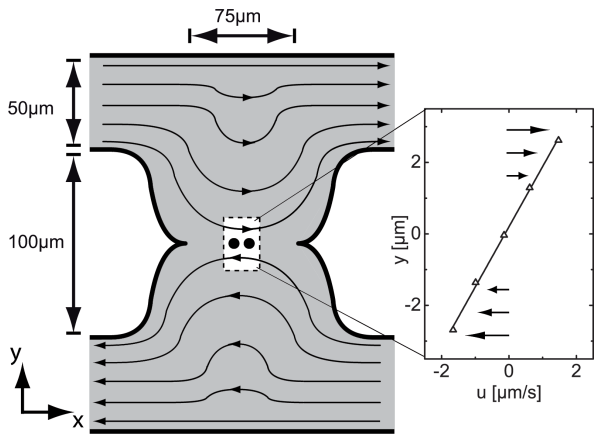, angle=0, width=0.9\linewidth}
\caption{A cell cross-section (lithographic mask) is shown with $150 \mu m$  depth
in the $z$ direction having opposite flow directions in its upper and lower channel
and a linear shear profile at its
center: see inset with PIV data ($\Delta$) and linear fit. 
One or two particles at a distance $d=4\mu m$ were held by 
optical tweezers in the center of the 
 linear velocity profile ${\bf u}(y)$.} \label{fig01}
\end{figure}

By a dual beam optical tweezer-setup, composed of two solid state lasers   and
an oil immersion objective with a numerical aperture of 1.4,
two harmonic potentials are generated 
in an inverted microscope (Nikon TE 2000-S) 
to capture uncharged polystyrene beads (Duke Scientific Corporation, R0300)
with a diameter of $3\mu m$ in a
flow of distilled water. 
The beads were observed
with a high speed camera (IDT, X-Stream, XS-5) of 15 kHz and their positions were
determined with a correlation tracking algorithm with a 
spatial resolution of $\pm 4$ nm \cite{measureII}. To avoid 
any interference of the two potentials at small distances and to
maximize the hydrodynamic interaction effect between
the two particles, they where held at a distance of $d=4 \mu m$.
In a microfluidic device with two counter flows, as shown in Fig.~\ref{fig01},
a linear shear gradient with a vanishing mean velocity was 
generated at the center of the cell, as experimentally verified by micro-PIV  (see
inset in Fig.~\ref{fig01}).
The design of the flow chamber was optimized by numerical simulations of the incompressible Navier-Stokes equation (Multiphysics 3.4, Comsol AB, Stockholm, Sweden). The small width in $x$ direction
of the center piece of the chamber was
chosen in order to minimize flow in $y$ direction. The curved
form of the boundaries was found to suppress vortices.
The channel was manufactured by standard soft lithographic techniques and the flow was driven by gravitational potential difference. 
 The distance of the beads from the wall was always larger than $10 \mu m$ in the $z$ direction
and $25 \mu m$ in the $xy$ plane.  Hence 
boundary effects on the bead fluctuations
could be excluded within our experimental resolution, as verified 
by measurements 
without flow, which were in very good agreement with
previous results (see e.g. Ref. \cite{Quake:99.1}). However, with flow
the experimental noise becomes larger,
especially at longer correlation times.
The value of the shear rate $\dot\gamma$ has been 
extracted from the fits of the correlation data and
by particle tracking methods. The highest shear rate that did 
not lead to an escape of the particles from the
traps was $ \dot \gamma \simeq 50 s^{-1}$ \cite{EPAPS}.

\begin{figure}[htb]
\epsfig{file=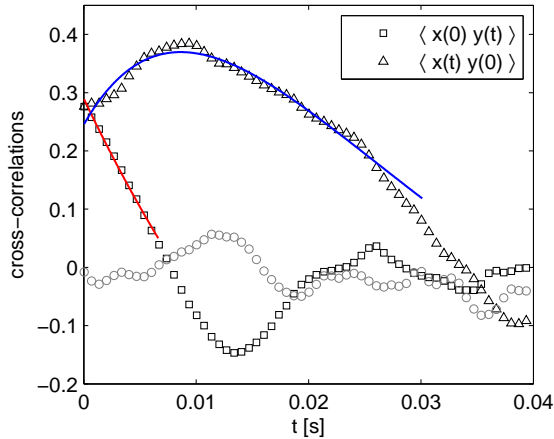, angle=0, width=0.9\linewidth}
\vspace{-2mm}
\caption{Shear-induced cross-correlations 
between orthogonal random displacements: $\langle x(t) y(0)\rangle$ (triangles) 
and $\langle x(0)y(t) \rangle$ (squares).
Lines are fits 
according to Eq.~(\ref{corrsingpxty}) and Eq.~(\ref{singlecorr})
and open circles represent $\langle x(0) y(t) \rangle$ 
in the absence of flow.
} 
\label{fig03}
\end{figure}

One or two Brownian particles with coordinates $\vecd{r}_i=(x_i,y_i,z_i)$\, ($i=1,2$) are held in a linear shear flow $\vecd{u}({y}_i)=\dot{\gamma}y_i\hat{\vecd{e}}_x$  
by forces ${\bf f}_i^V=k ({\bf p}_i-{\bf r}_i)$ close
to the minima ${\bf p}_i$   of two harmonic potentials 
$V_i=\frac{k}{2}(\vecd{p}_i-\vecd{r}_i)^2$ (spring constant $k$).
The over-damped particle motion 
is described by a Langevin equation \cite{Dhont:96}:
\begin{align}
 \label{dgl1}
 \dot{\vecd{r}}_i =\vecd{u}(\vecd{r}_i) + {\mat{H}}_{ij}\left( \vecd{f}^V_j + \vecd{f}^S_j \right)\,.
\end{align}
The mobility matrix ${\mat{H}}_{ij}$ accounts for the Stokes friction
 and the hydrodynamic interactions between them. Here we
use the Oseen approximation
\begin{align}
  {\mat{H}}_{11}&=\mat{H}_{22}=\frac{1}{\zeta}\mat{E}\,,\\
  {\mat{H}}_{12}&={\mat{H}}_{21}= \frac{1}{\zeta} \frac{3a}{4 r_{12}} \left[{\mat{E}}+\frac{\vecd{r}_{12}\vecd{r}_{12}^T}{r_{12}^2}\right]\,,
\end{align}
with the Stokes friction coefficient
$\zeta= 6 \pi \eta a$ of a point particle 
of effective hydrodynamic
radius $a$ in a fluid
of viscosity $\eta$ and the unity matrix 
$\mat{E}$. $\vecd{r}_{12}=\vecd{r}_1-\vecd{r}_2$ is the bead distance and $r_{12}$ is its norm, $\tau= \zeta/k$ the particle relaxation time in the potential and $W = \dot{\gamma} \tau$ the Weissenberg number.
The Brownian particle motion
is driven by the stochastic forces $\vecd{f}^S_i(t)$ in Eq.~(\ref{dgl1}), for which we assume vanishing
mean values and correlation times:
\begin{align}
 \label{stoch_eq1}
 \langle \vecd{f}^S_i(t) \rangle &= 0\,, \\
 \label{stoch_eq2}
 \langle \vecd{f}^S_i(t)  \vecd{f}^S_j(t') \rangle &= 2k_BT {\mat{H}}^{-1}_{ij} \delta(t-t')  \,.
\end{align}
At first we investigate the Brownian motion of
a single trapped particle in shear flow. Its autocorrelation 
along  the flow direction, 
 $\langle x(0)x(0) \rangle = \frac{k_B T}{k}( 1 + W^2/2)$, depends on $W$, 
 but along the perpendicular direction,
$\langle y(t)y(0)\rangle = \frac{k_BT}{k}~\exp\left(-\frac{t}{\tau}\right)$\,,
it does not depend on $W$  \cite{HolzBamm}.
In a quiescent fluid 
cross-correlations between particle displacements in {\it orthogonal
directions} vanish:
$\langle x(t)y(t') \rangle=0$.  But shear flow 
causes in the shear plane finite cross-correlations \cite{Bedeaux:1995.1,Rzehak:2003.2,Reeks:2005.1}, which are
 asymmetric with respect to $t \to -t$ 
\cite{HolzBamm}:
\begin{eqnarray}
\label{corrsingpxty}
\langle x(t)y(0) \rangle &=&\frac{k_B T}{k} \frac{W}{2}~ e^{ -t/\tau}
 \left(1+2 \frac{t}{\tau} \right)\,,\\ 
\label{corrsingpxyt}
\langle x(0)y(t) \rangle &=&\frac{k_B T}{k} \frac{W}{2} ~ e^{ -t/\tau}.
\label{singlecorr}
\end{eqnarray}
The algebraic prefactor in Eq.~(\ref{corrsingpxty}) illustrates that 
a fluctuation $y(0)\not =0$ of a particle is carried away
by the flow in the $x$ direction before the initial 
displacement  $y(0)$ relaxes. This leads,
during an initial period shorter than the relaxation time $\tau$,
to a growth of $\langle x(t)y(0) \rangle $, while the expression in Eq.~(\ref{singlecorr})
decays monotonically. As shown
in Fig.~\ref{fig03}, the predicted elementary signatures for
the shear-induced cross-correlations, cf. Eq.~(\ref{corrsingpxty}),
are in agreement with
our experimental data (triangles). 
Here $\langle x(t)y(0) \rangle $ takes its maximum roughly at $t\approx 0.009 s$,
corresponding via Eq.~(\ref{corrsingpxty}) to a particle's 
relaxation time $\tau \approx 0.018 s$.
Also the initial decay of $\langle x(0) y(t) \rangle$ 
(squares in Fig.~\ref{fig03}) agrees with 
our model, cf. Eq.~(\ref{corrsingpxyt}). The additionally observed 
minimum is possibly caused by a slight inclination of the laser beam or 
it is a reminiscent of a long wave length oscillation due to
the limited number of samples taken \cite{EPAPS}.
 For fluid-velocity fluctuations in orthogonal directions 
in the shear plane a similar signature as in Eq.~(\ref{corrsingpxty}) has been found
 \cite{Eckhardt:2003.1}.
According to Eq.~(\ref{corrsingpxty} and Eq.~\ref{singlecorr}) one obtains
the normalized ratios of the static cross-correlations: $\langle x(0) y(0) \rangle/ \langle y(0) y(0) \rangle = W/2$ and $\langle x(0) y(0) \rangle/ \langle x(0) x(0) \rangle = \frac{W/2}{1+W^2/2}$ \cite{HolzBamm}. 
From the fits, as indicated by the
red and blue line  in  Fig.~\ref{fig03},
we obtain $\langle x(0) y(0) \rangle/\langle x(0)x(0) \rangle \approx 0.26$, 
which corresponds to a Weissenberg number $W\approx 0.62$.

\begin{figure}[htb]
\epsfig{file=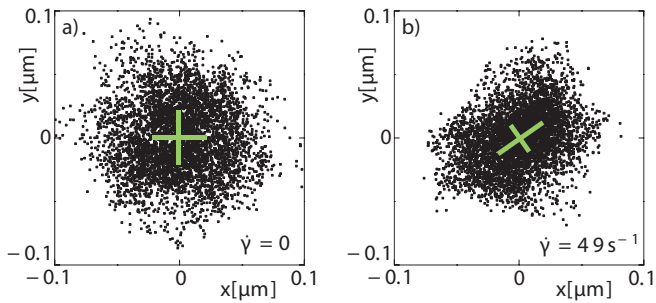,angle=0, width=1.0\linewidth}
\caption{The particle's distribution in the shear plane is shown in a) 
without flow and for a shear flow with $\dot \gamma = 49 s^{-1}$ in b). The angle
between the major and the $x$ axis is $\phi \approx 38^o$ 
and the ratio between the two principal axes is $R \approx 0.75$.}
\label{fig02}
\vspace{-0.0cm}
\end{figure}

The probability distribution of a Brownian particle in 
a harmonic potential and exposed to a linear shear flow
has an elliptical cross-section
as shown by the particle's position in Fig.~\ref{fig02} but it
has circular symmetry in the
absence of flow.
The angle $\phi$ enclosed by the major axis of the particle's probability
 distribution and
the $x$ axis, as well as the ratio $R$ between the lengths of the two 
principal axes, in the shear plane, depend on the Weissenberg number $W$ as follows \cite{HolzBamm}:
\begin{eqnarray}
 \tan \phi &=& \frac{1}{2}
\left[ \sqrt{4+W^2}  - W \right]\,,
\label{tanp}\\
 R&=&\left(\frac{\sqrt{4+W^2}-W}{\sqrt{4+W^2}+W}\right)^{1/2}\,.
\label{V}
\end{eqnarray}
Using $W \approx 0.62$ as determined above,
 one obtains via  Eq.~(\ref{tanp}) the
angle $\phi \approx 37^o$
and via Eq.~(\ref{V}) the ratio 
$R \approx 0.72$.  Within 
errors this is consistent with the angle 
 $\phi \approx 38^o$ and
the ratio $R \approx  0.75$  obtained
from the measured particle's distribution
shown in Fig.~\ref{fig02}.

For two particles, each trapped in a potential minimum  in shear flow, we investigated
the correlations between their 
random displacements for two different configurations: with
 the connection vector ${\bf p}_{12}={\bf p}_1-{\bf p}_2$ 
parallel to the flow direction as in Fig.~\ref{fig01} or perpendicular to it.

\begin{figure}[htb]
\epsfig{file=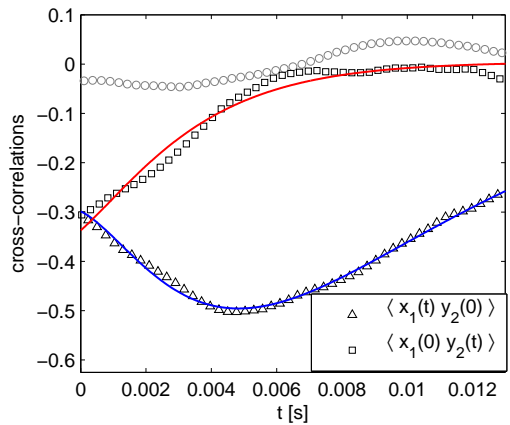, angle=0, width=0.85\linewidth}
\caption{Correlations $\langle x_1(t) y_2(0)\rangle$ (triangles) and $\langle x_1(0) y_2(t) \rangle$ (squares) between random displacements
of two particles. Colored lines are fits according to Eq.~(\ref{parx0yt}) and Eq.~(\ref{parxty0}). Circles represent the same correlations
in the absence of flow.}
\label{fig04}
\vspace{-1mm}
\end{figure}

For Brownian displacements of the two distinct particles
along the same direction the quantities $\langle x_i(t), x_j(0)\rangle$ and
$\langle y_i(t), y_j(0) \rangle $ describe
anti-correlations
for $i\not =j$ (see e.g. Ref.~\cite{Quake:99.1}). The
shear-induced corrections for both are of the order of $W^2$ as described in
more detail in  Ref.~\cite{HolzBamm}. For random displacements of distinct
particles, but along orthogonal directions, one only finds correlations
in the presence of shear flow.  With the abbreviations
\begin{eqnarray}
 \lambda_{1,3}=1\pm 2 \mu\,,~~ \lambda_{2,4}=1\pm \mu\,,~~ \mu= \frac{3 a}{4d}\,,
\end{eqnarray} 
and the connection vector ${\bf p}_{12}$ parallel to the
flow two of the anti cross-correlations 
in the shear plane are \cite{HolzBamm}
\begin{eqnarray}
&&\hspace{-4mm}\langle x_1(t)y_2(0) \rangle =\frac{k_B T}{\mu k} \frac{W}{2}
\left(  e^{-\lambda_2t/\tau }+e^{-\lambda_4 t/\tau} \right. \nonumber \\
&& \hspace{1.5cm} - \left. \frac{2\lambda_2e^{-\lambda_1 t/\tau} }{2+3\mu} -  \frac{2\lambda_4e^{-\lambda_3 t/\tau }}{2-3\mu} 
\right)\,, 
\label{parx0yt}\\
&&\hspace{-4mm}\langle x_1(0)y_2(t) \rangle =\frac{k_B T}{k} \frac{W}{2} \left( \frac{e^{-\lambda_2 t/\tau}}{2+3\mu} -  \frac{e^{-\lambda_4 t/\tau}}{2-3\mu}\right)\,.
\label{parxty0}
\end{eqnarray}
The cross-correlation $\langle x_1(t)y_2(0) \rangle $
(triangles) in Fig.~\ref{fig04} and the fit (blue line)
show a pronounced minimum at about the particles
relaxation time $t \approx \tau$. 

With a connection vector ${\bf p}_{12}$ 
perpendicular to the flow lines we obtain
a cross-correlation $\langle x_2(t) y_1(0)$ in
the limit of small values of $\mu$,
\begin{eqnarray}
\langle x_2(t)y_1(0) \rangle \approx - \frac{k_B T}{2k} \frac{W\mu}{2} ~e^{-t/\tau} \left(3 + 
2\frac{t}{\tau}+6 \frac{t^2}{\tau^2}  \right)\,,\nonumber
\end{eqnarray}
which exhibits in contrast to Eq.~(\ref{parx0yt}) two extrema.

Shear-induced  cross-correlations between random displacements of a 
single particle in a potential were calculated and measured here 
for the first time, to the best of our knowledge, cf. Fig.~\ref{fig03}. At approximately
half of the particle's relaxation time $\tau$ the correlation function in
Eq.~(\ref{corrsingpxty}) exhibits with its maximum a
typical signature of Brownian motion in shear flow, caused
 by the rotational part of the shear flow as well as
the non-normal property of the linearized Navier-Stokes equation.
Simultaneously, for a particle in a harmonic potential and shear flow
 an elliptical probability distribution was measured. Both independent measurements 
are described by a Langevin model for the same value of the
 Weissenberg number, which confirms the validity of our
approach to shear-flow effects on
the Brownian particle dynamics.

Theoretically, shear-induced correlations between perpendicular
fluid-velocity fluctuations have been investigated before \cite{Eckhardt:2003.1,Oberlack:2006.1}. 
Those are traced back to 
the non-normal property of the linearized Navier-Stokes
equation \cite{Eckhardt:2003.1} and they are
important for the stability of shear flow and the
onset of turbulence.  The cross-correlations between these
 velocity fluctuations are based on the
same mechanism as discussed here and  they exhibit  similar extrema
as our experimental and analytical results.

Stochastic forces on a suspended particle are caused by 
velocity fluctuations of the surrounding fluid. Usually, they are assumed to be 
isotropic in related Langevin models with
uncorrelated perpendicular components. However,
cross-correlations of the velocity fluctuations 
in shear flow,
as discussed  in Refs.~\cite{Eckhardt:2003.1,Oberlack:2006.1}, 
will modify the cross-correlations
between orthogonal particle displacements, as investigated
here, but the related additional contributions 
to the particle displacement correlations 
are expected to be considerably smaller than the effects
of isotropic random forces \cite{HolzBamm}.
It is, however, an interesting and challenging future
issue to separate these two non-equilibrium effects 
in experiments.

For two hydrodynamically interacting particles, 
each captured by an optical tweezer
at the center of the shear flow,
we find shear-induced anti-correlations 
between orthogonal particle displacements with one extremum
if the vector connecting the mean particle positions
is parallel to the streamlines and two extrema, 
if the connection vector is perpendicular to the flow lines.
These properties may be relevant for further understanding of
the dynamics of polymer models in shear flow.

This work was supported by the German science foundation via 
the priority program on micro- and nanofluidics SPP 1164 
and the graduate school 
GRK 1276.

\end{document}